\documentstyle[12pt,aaspp,flushrt]{article}
\newcommand{\Ref}{\hangindent=20pt \hangafter=1 \noindent}
\newcommand{\StartRef}{\hyphenpenalty=10000 \raggedright}

\newcommand{\beq}{\begin{equation}}
\newcommand{\eeq}{\end{equation}}
\newcommand{\NarrowMargins}{
  \setlength{\oddsidemargin}{+0.3in}
  \setlength{\evensidemargin}{-0.0in}
  \setlength{\textwidth}{6.2in}
  \setlength{\topmargin}{-0.75in}
  \setlength{\textheight}{9.25in}   }
\catcode`\@=11 
\def\xmp{{x_M^{\prime}}}
\def\lsim{\mathrel{\mathpalette\@versim<}}
\def\gsim{\mathrel{\mathpalette\@versim>}}
\def\@versim#1#2{\vcenter{\offinterlineskip
        \ialign{$\m@th#1\hfil##\hfil$\crcr#2\crcr\sim\crcr } }}
\catcode`\@=12 
\NarrowMargins
\begin{document}
\title{Are Particles in  Advection--Dominated Accretion Flows Thermal?} 
\author{Rohan Mahadevan\footnote{rmahadevan@cfa.harvard.edu} and  
Eliot Quataert\footnote{equataert@cfa.harvard.edu}}
\affil{Harvard-Smithsonian Center for Astrophysics,
60 Garden St., Cambridge, MA 02138}
\setcounter{footnote}{0}
\begin{abstract}
We investigate the form of the momentum distribution function for
protons and electrons in an advection--dominated accretion flow
(ADAF).  We show that for all accretion rates, Coulomb collisions are
too inefficient to thermalize the protons.  The proton distribution
function is therefore determined by the viscous heating mechanism,
which is unknown.  The electrons, however, can exchange energy quite
efficiently through Coulomb collisions and the emission and absorption
of synchrotron photons. We find that for accretion rates greater than
$\sim 10^{-3}$ of the Eddington accretion rate, the 
electrons have a thermal distribution throughout the accretion flow.

For lower accretion rates, the electron distribution function is
determined by the electron's source of heating, which is primarily adiabatic
compression.  Using the principle of adiabatic invariance, we show
that an adiabatically compressed collisionless gas maintains
a thermal distribution until the particle energies become
relativistic.  We derive a new, non--thermal, distribution function
which arises for relativistic energies and provide analytic
formulae for the synchrotron radiation from this distribution.
Finally, we discuss its implications for the emission
spectra from ADAFs. \\

\noindent {\em Subject headings:} accretion, accretion disks --- 
black hole physics --- radiation mechanisms: thermal, non--thermal 
--- galaxies: elliptical and lenticular, cD --- radio continuum: galaxies
--- Galaxy: general
\end{abstract}
\section{Introduction}
An advection--dominated accretion flow is a hot, optically
thin, accretion flow with low radiative efficiency (Abramowicz et
al. 1988; Narayan \& Yi 1994, 1995a, 1995b; Abramowicz et al. 1995).
Unlike standard thin disks (see Frank et al. 1992), where all of the
viscously generated energy is thermalized and radiated locally, ADAFs
store most of the viscously generated energy as internal energy of the
ions and advect it onto the central object.  The gas in an ADAF is a
two temperature plasma, with the ions being significantly hotter than
the electrons (Shapiro et al. 1976; Rees et al. 1982).  Since most of
the viscously generated energy is assumed to heat the ions, and only a
small fraction of this energy is transferred to the electrons via
Coulomb collisions, the total energy radiated is much less than the
total energy generated by viscosity (Rees et al. 1982).

The emission spectrum from an ADAF is determined by the cooling
mechanisms of the protons and electrons.  The protons cool very
inefficiently, primarily through proton--proton collisions which
create neutral pions.  The pions decay into $\gamma$--rays with mean
frequencies $\sim 10^{24}$ Hz (Mahadevan et al. 1997).  The electrons,
on the other hand, cool very efficiently through synchrotron,
bremsstrahlung, and Compton processes.  Detailed calculations
of these cooling mechanisms lead to ADAF models which have
been successfully applied to explain the observed radio to gamma-ray
spectrum from a number of accreting black hole systems.  These include
solar mass black holes in X-ray binaries, e.g., A0620 and V404Cyg
(Narayan, McClintock, \& Yi 1996), as well as supermassive black holes
at the centers of galaxies, e.g., Sgr A$^*$ (Narayan et al. 1995;
Mahadevan et al. 1997) and NGC 4258 (Lasota et al. 1996).

The shape and magnitude of an ADAF spectrum depends primarily on the
mass accretion rate and, to a lesser extent, on the mass of the
accreting object (Narayan \& Yi 1995b; Narayan 1996).  In determining
the spectrum, however, all previous work has assumed that
the electrons are thermal throughout the ADAF.  This is a crucial
assumption, since the emergent spectrum would be considerably
different for different electron distribution functions.  In the
present paper, we investigate the validity of this assumption.

We treat thermalization within the ADAF paradigm.  We assume that the
only physical interactions among particles are Coulomb collisions and
self--absorbed synchrotron radiation.  We do not consider the possible
non--thermal heating of electrons by wave--particle
interactions, magnetic reconnection, or collective plasma effects.
The interactions considered here therefore incorporate only the well
understood physical processes in ADAFs.

The outline of the paper is as follows. In the following section (\S
2) we present the necessary background equations and consider proton
thermalization by Coulomb collisions.  In \S 3 we investigate the form
of the electron distribution function, in particular for suprathermal
electron energies, and determine the mass accretion rates for which  
electrons are thermalized by Coulomb collisions or self--absorbed
synchrotron radiation.  In \S 4 we consider the form of the electron
distribution function when thermalization is inefficient, which
depends on the interplay between the heating and cooling mechanisms.
We give a new (non-thermal) distribution function, valid when the
electrons compress adiabatically to relativistic energies.  In
\S 5 we give the synchrotron spectrum from this new distribution
function and in \S 6 we discuss the implications and  possible 
applications of our results.
\section{General Relations}
\subsection{Self-similar Flow Equations}
The parameters which determine the structure of an ADAF are the
viscosity parameter, $\alpha$ (Shakura \& Sunyaev 1973), the ratio of
the gas pressure to the total pressure, $\beta_{\rm adv}$, the mass of
the central object, $M = m \ M_{\odot}$, where $m$ is the mass in
solar mass units, and the accretion rate, $\dot{M} =
\dot{m} \ \dot{M}_{\rm Edd}$, where $\dot{m} $ is the accretion rate in Eddington 
units ($\dot{M}_{\rm Edd} = 1.39\times 10^{18} m$ g s$^{-1}$).  Typical 
values for $\alpha$ and $\beta_{\rm adv}$ which have been successfully
applied to observed systems are $0.3$ and $0.5$, respectively 
(Narayan et al. 1996, 1997).  $\beta_{\rm adv} = 0.5$ corresponds to 
equipartition between magnetic and gas pressure, as suggested by
the work of Balbus and Hawley (1991).

Unlike standard thin accretion disks, an ADAF is well approximated by
a series of concentric spherical shells with the properties of the gas
varying as a function of radius (Narayan \& Yi 1995a).  The self
similar solution of Narayan and Yi (1994, 1995a) provides reasonably
accurate analytical estimates of the properties of the accretion flow.
For the present discussion, the quantities of interest are
$${v(r) \over c } \simeq  0.37 \, {\alpha}  \, r^{-1/2}, $$
$$\theta_p(r) \simeq 0.09 \left( {\beta_{\rm adv} \over 0.5} \right) r^{-1},$$
$$ n_e(r) \simeq 6.3 \times 10^{19} \, 
		{\alpha^{-1} }  \, m^{-1} \dot{m} \, 
		r^{-3/2} \ \ \mbox{cm$^{-3}$},  $$ 
\beq
B = 7.7 \times 10^8 \, \alpha^{-1/2} 
		\left( {1-\beta_{\rm adv} \over 0.5} \right)^{1/2} \, 
	m^{-1/2}\dot{m}^{1/2} \, r^{-5/4} \ \ \mbox{Gauss},  \label{propeq}
\eeq
where $v(r)/c $ is the radial velocity in units of the velocity of light, 
$\theta_p(r) = kT_p/m_p c^2$ is the dimensionless proton temperature,
$n_e(r)$ is the number density of electrons, $B(r)$ is the magnetic field
strength and $r = R/R_S$ is the radius in Schwarzschild units ($R_S =
2.95 \times 10^5 m$ cm).  In equation (\ref{propeq}),
the fraction, $f$, of the viscously dissipated energy that is carried inward by 
the accreting gas is taken to be $\sim 1$ (Narayan et al. 1996).

\subsection{Timescales}
For the electrons or protons in an ADAF to remain thermal as they
accrete, they must have sufficient time to redistribute their kinetic
energy by various processes.  We therefore take thermalization by
Coulomb collisions to be efficient when the timescale for collisions
between the electrons, $t_{\rm ee}$, or between the protons, $t_{\rm
pp}$, is shorter than the accretion time, $t_{\rm a}$.  Since the
protons are non-relativistic ($\theta_p < 0.1$) their thermalization
time is (Spitzer, 1962)
\begin{eqnarray}
t_{\rm pp}  &=& {(2\pi)^{1/2} \over n_p \sigma_T c \ln \Lambda} \, 
	\left( {m_p  \over m_e} \right)^2 \, \theta_p^{3/2}, \nonumber \\
&\simeq& 9.2 \times 10^{-3}  {\alpha } \, 
	\left( {\beta_{\rm adv} \over 0.5}\right)^{3/2} \, m \, \dot{m}^{-1} 
\ \ \ \mbox{s}, \label{tpp}
\end{eqnarray}
where $n_p = n_e$ is the proton number density, $\sigma_T$ is the
Thomson cross--section, $\ln \Lambda \approx 20$ is the Coulomb
logarithm, and we have used equation (\ref{propeq}) for the proton
temperature profile.  
The time required for the protons and electrons to reach
thermal equilibrium is (Spitzer 1962)
\begin{eqnarray}
t_{\rm ep} &=& 
{(2\pi)^{1/2} \over 2 \, n_e \sigma_T c \ln \Lambda} \, 
	\left( {m_p  \over m_e} \right) \, (\theta_e + \theta_p)^{3/2}, \nonumber \\
&\simeq& 9.3 \times 10^{-5}  {\alpha } \, 
	\theta_e^{3/2} \, m \, \dot{m}^{-1} \, r^{3/2} \ \ \ {\rm s}. \label{tep}
\end{eqnarray}
So long as the electrons are non-relativistic
their timescale for thermalization is  (Spitzer 1962)	
\begin{eqnarray}
t_{\rm ee} &=& {(2 \pi)^{1/2} \over n_e \sigma_T c \, \ln \Lambda} \
\theta_e^{3/2}, \nonumber \\ &\simeq& 1.0\times 10^{-7} \, {\alpha} \,
m \, \dot{m}^{-1} \, r^{3/2} \
\theta_e^{3/2} \ \ \ {\rm s}.
\label{tee}
\end{eqnarray}

For thermalization to occur before the gas falls into the central
black hole, the above timescales must be less than the accretion time of
the gas, which is
\beq
t_{\rm a} = \int {dR \over v(R)} \simeq 1.8\times 10^{-5} \, 
{\alpha}^{-1} \, m \, r^{3/2} \ \ \ \mbox{ s}.
\label{ta}
\eeq

From these timescales we can readily estimate the critical accretion
rate, $\dot{m}_{\rm crit}$, above which the ADAF solution ceases to
exist.  For sufficiently large mass accretion rates, the electrons and
protons become thermally well coupled and so the assumption of a two
temperature plasma is no longer valid; the two temperature ADAF therefore  
no longer exists.  The accretion rate above which this occurs is
obtained by setting $t_{\rm ep} \sim t_{\rm a}$, which gives
\beq 
\dot{m}_{\rm crit} \approx  0.45 \, \alpha^2, \label{mdotcrit}
\eeq 
where we have set $\theta_e \sim 0.2$, which is valid for high
$\dot{m}$ systems (Mahadevan 1997).  This is comparable to the
critical accretion rates obtained by independent arguments (Narayan \&
Yi 1995b; see also Mahadevan 1997).

\subsection{Proton Thermalization}

To determine the accretion rates at which the protons have sufficient
time to thermalize by Coulomb collisions, we set $t_{\rm pp} \lsim
t_{\rm a}$. This gives the requirement
\beq
\dot{m} \gsim  500 \, \alpha^2 \, \left({\beta_{\rm adv} \over 0.5}\right)^{3/2} r^{-3/2}.
\eeq
Since the maximal accretion rate for an ADAF is $\dot{m}_{\rm crit}$,
the radius above which the protons can thermalize is
\beq
r_{\rm th} \approx 100 \, \left({\beta_{\rm adv} \over 0.5} \right)
\left( {\dot{m} \over \dot{m}_{\rm crit}}
\right)^{-2/3}. 
\label{rmin}
\eeq
For a given $\dot{m}$, the protons are unable to redistribute their
energy through Coulomb collisions for $r < r_{\rm th}$ and so the
proton distribution function at these radii is primarily determined by
the viscous heating mechanism.  Observational probes of the proton
distribution function would therefore provide valuable information about
the details of the viscous heating mechanism. Mahadevan et al. (1997)
have shown that the $\gamma$--ray spectrum from pions created by proton collisions in
ADAFs can be used as a direct probe of the proton
distribution function, and can therefore provide a means of studying
viscosity in hot accretion flows.
\section{The Electron Distribution Function in ADAFs}
In this section we analyze the form of the electron distribution
function in ADAFs.  We show that for moderately high accretion rates,
Coulomb collisions and synchrotron self--absorption can efficiently
thermalize the electrons, even in the high energy tail of the
distribution function.  At lower accretion rates, however, these
processes are too inefficient, and the form of the distribution
function depends on the dominant heating and cooling mechanisms of the electrons.
This is discussed in \S4.
\subsection{Coulomb Collisions}
The accretion rate, $\dot m_{\rm th}$,  above which the electrons can thermalize via
Coulomb collisions is obtained by setting the
timescale for electron--electron collisions, $t_{\rm ee}$, equal to
the accretion time, $t_{\rm a}$, which gives
\beq
\dot{m}_{\rm th} \approx 3.8\times 10^{-4} \, T_9^{3/2}  \, \alpha^2,
\label{mdth}
\eeq
where $T_9$ is the electron temperature in units of $10^9$ K.  For
$\dot{m} \gsim \dot{m}_{\rm th}$, electrons with energies $\sim kT_e$
are thermal.  Since most of the synchrotron emission from ADAFs comes from the
high energy tail of the distribution (\S 4), it is also necessary to
determine whether electrons with energies $\gg kT_e$ have enough time
to thermalize.  To investigate this, we use the Fokker--Planck
equation which, for an isotropic energy distribution function, takes
the form
\beq
{\partial n(\gamma,t) \over \partial t} = 
- { \partial\over \partial \gamma}\, \left[ \left( {d \gamma \over d t} \right)
\, n(\gamma,t)\right] + {1 \over 2} \, {\partial ^2 \over \partial \gamma^2} \, 
\left[ {d(\Delta\gamma)^2 \over d t} \, n(\gamma, t) \right],
\label{fpeq}
\eeq
where $n(\gamma,t)$ is the distribution function, $d\gamma/dt$
represents systematic acceleration, and $d(\Delta\gamma)^2 / d t$
represents stochastic acceleration.  The systematic acceleration is
negative for particles with $\gamma \gsim \theta_e$, and positive for
$\gamma \lsim \theta_e$, which corresponds to slowing down (speeding
up) particles with energies large (small) compared to the mean energy
of the plasma.  The stochastic acceleration 
is always positive and corresponds to diffusion in energy space.

The timescale for an electron with a Lorentz factor $\gamma$ to
thermalize with a background plasma of temperature $\theta_e$ is given
roughly by the timescale on which the electron's energy e-folds.  This
ensures that the electron interacts significantly with the plasma. The
resulting timescale depends on $\gamma$ and $\theta_e$, and is
obtained by determining whether systematic or stochastic acceleration
dominates the energy changes of the electron.

The systematic acceleration of an electron with Lorentz factor $\gamma$, 
interacting with a thermal plasma of temperature $\theta_e$, is given
approximately by (Dermer \& Liang 1989)
\beq
{d \gamma \over d t} = - \, {3 \over 2} \, n_e \sigma_T c \ln \Lambda \,
\left[ {1\over A(\theta_e)} - {1 \over \gamma} \right],
\nonumber \\
\label{dldirect} 
\eeq
where $A(\theta_e) = K_2(1/\theta_e) / K_1(1/\theta_e)$ and $K_n$ is
the modified Bessel function of order $n$. The timescale on which the
electron's energy e--folds is therefore
\beq
t^{\rm e}_{\rm sys} = {1 \over |d\ln \gamma / dt|} =  
	{2 \over 3} \, {\gamma \, \over  n_e \sigma_T c \ln \Lambda} \, 
	\left| {1\over A(\theta_e)} - 
{1 \over \gamma} \right|^{-1}.
\label{directc}
\eeq
Equations (\ref{dldirect}) and (\ref{directc}) are good analytical
approximations to detailed numerical calculations provided that
$\theta_e \gsim 0.3$ and $\gamma \gsim 2$ (Dermer \& Liang 1989).

The stochastic acceleration is determined by substituting equation
(\ref{dldirect}) into the Fokker-Planck equation, setting $\partial
n/\partial t = 0$ (as is required for a thermal plasma) and solving
the second order differential equation.  The two integration constants
in the resulting solution are determined by (1) requiring no net flux
of electrons in energy space (since there are no sinks or sources of
electrons) and (2) imposing the physical requirement that the
stochastic acceleration not increase exponentially for large $\gamma$.
This gives
\beq
{d(\Delta \gamma)^2 \over d t } = 
3 \, n_e \sigma_T c \ln \Lambda \, \left[ 
{\theta_e \over A(\theta_e)}
+ {\theta_e \over \gamma } z_{\theta_e} + {\theta_e^2 \over \gamma^2} z_{\theta_e}
		\right],
\label{neweq}
\eeq
where $z_{\theta_e} = 2[\theta_e/A(\theta_e)]-1 $. The resulting diffusion time is
\beq
t^{\rm e}_{\rm diff} = 1\left/ {1 \over \gamma^2} \,{d(\Delta \gamma)^2 \over d t} \right.
 =  {\gamma^2 \over 3 \, n_e \sigma_T c \ln \Lambda} \, 
 \left[{\theta_e \over A(\theta_e)} + {\theta_e \over \gamma }z_{\theta_e} + 
{\theta_e^2 \over \gamma^2} z_{\theta_e}
		\right]^{-1}. 
\label{diffc}
\eeq
Since equations (\ref{neweq}) and (\ref{diffc}) are derived from equation 
(\ref{dldirect}), they are also only valid for $\theta_e \gsim 0.3$ and $\gamma
\gsim 2$. We note that the stochastic acceleration derived here does not 
rise exponentially as
indicated in Dermer \& Liang (1989; cf. their eq. [14] and figure 2).  More recent 
detailed calculations by Nayakshin \& Melia (1997) agree with the analytic 
approximation for the stochastic acceleration given here.

Setting $(t_{\rm sys}, t_{\rm diff}) < t_{\rm a}$ and using equations 
(\ref{propeq}) and (\ref{ta}) gives the accretion
rates required for thermalization,
\begin{eqnarray}
\left({\dot{m} \over \alpha^2} \right)^{\rm e}_{\rm sys} &\gsim& 
1.4\times 10^{-3} \, \gamma \,  \left| { 1 \over A(\theta_e)} -
	{1 \over \gamma} \right|^{-1}, \nonumber \\
\left({\dot{m} \over \alpha^2} \right)^{\rm e}_{\rm diff} &\gsim& 
7.3 \times 10^{-4} \, \gamma^2 \, 
\left[{\theta_e \over A(\theta_e)} + {\theta_e \over \gamma }z_{\theta_e} +
	{\theta_e^2 \over \gamma^2}z_{\theta_e} \right]^{-1}.
\label{eemdot}
\end{eqnarray}
For electrons with energies comparable to the mean energy of the
plasma, $\gamma \sim \theta_e$ and equation (\ref{directc}) shows that
$t^{\rm e}_{\rm sys} \rightarrow \infty$; the electron, on average, is
not heated by direct acceleration, but instead diffuses in energy
space (as determined by the stochastic term).  For large
$\gamma$, $t^{\rm e}_{\rm sys} < t^{\rm e}_{\rm diff}$, and so electrons with
energies much larger than the mean energy of the plasma,  $\gamma
\gg \theta_e$, first lose energy by direct cooling and then relax 
by diffusion.  In order to assess the efficiency of
thermalization, we therefore take the minimum of the systematic and
diffusion timescales as the relevant timescale against which to
compare the accretion time.

For a plasma at temperature $\theta_e$, an electron with Lorentz
factor $\gamma$ can thermalize via Coulomb collisions if the accretion
rate is greater than the smaller of the two accretion rates given in
equation (\ref{eemdot}).  The dashed lines in figure 1 show this
accretion rate as a function of the Lorentz factor $\gamma$, for two
values of $\theta_e$.  For accretion rates below these curves, the
electrons do not have time to interact with one another via collisions
and are ``frozen'' into
their initial distribution.  
Unless other thermalization processes are important, this implies that
the form of the distribution function can deviate substantially from a
Maxwellian.  We show below that because the
synchrotron photons in ADAFs are highly self-absorbed, the electrons
are able to thermalize at lower accretion rates (and higher $\gamma$)
by interacting via the exchange of self--absorbed synchrotron photons.
\subsection{Electron Thermalization Through Synchrotron Self-Absorption}
Synchrotron self--absorption has been treated in the context of
finding steady state power--law electron distributions (McCray 1969),
heating low energy electrons through self--absorption (Ghisellini et
al. 1988), and as a thermalizing mechanism for low energy electrons
(Ghisellini \& Svensson 1990).  In particular, Ghisellini and Svensson (1990)
have shown that self-absorption of synchrotron photons leads to a
thermal distribution of electrons over a large range of $\gamma$ in a
few synchrotron cooling times.  The basic thermalization mechanism is
energy exchange via the absorption and emission of synchrotron
photons, which can be important even when the plasma is effectively
collisionless.

In keeping with the analysis of the previous section we discuss
thermalization induced by synchrotron self--absorption using the Fokker--Planck
equation.  Instead of attempting to solve the complete
integrodifferential equation (eq. [\ref{fpeq}]), we check for
consistency by assuming that the (thermal) synchrotron radiation is
highly self--absorbed, and then determining whether the thermalization
timescales obtained from the Fokker--Planck coefficients are faster
than the accretion time.

The kinetic equation governing the electron distribution function can
be written as a Fokker--Planck equation with coefficients given by
(McCray 1969; Ghisellini et al. 1988)
\beq
{d \gamma \over d t} =  - {1 \over m_e c^2} \int_0^{\infty} j(\nu, \gamma) \, d\nu + 
{1 \over \gamma \, p} {\partial \over \partial \gamma} \, \left[ 
	\gamma  p \, C(\gamma) \right], \label{synchdir}
\eeq
\beq
{d (\Delta\gamma)^2 \over d t} = 2 \,  C(\gamma), \\ \label{synchdiff}
\eeq
where 
\beq
C(\gamma)  \equiv {1 \over 2 m_e^2 c^2 } \int_0^{\infty} 
	{I(\nu) \over \nu^2} \, j(\nu, \gamma) \, d\nu. \label{synchceq}
\eeq
$j(\nu, \gamma)$  is the cyclo--synchrotron emissivity 
(ergs s$^{-1}$ Hz$^{-1}$),
$p = \gamma \, \beta = (\gamma^2 -1)^{1/2}$ 
is the electron's momentum in units of $m_e c$, and $I(\nu)$ is the
intensity of the background photons (ergs cm$^{-2}$ s$^{-1}$ Hz$^{-1}$).  

To evaluate $C(\gamma)$, we use three properties of synchrotron
radiation in ADAFs. First, the electrons which are responsible for
most of the synchrotron emission are from the high energy tail of the
distribution.  By multiplying the relativistic
Maxwell--Boltzmann distribution by the synchrotron emissivity and
using the method of steepest descent, we can determine the Lorentz factor
($\gamma_{\rm max}$) of the electron responsible for most of the
emission. This gives
\beq
\gamma_{\rm max}  =  2^{1/3} \theta_e x_M^{1/3} \sim 10 \theta_e,
\label{gammamaxeq}
\eeq
where $x_M \sim 500$ and is a very weak function of $\dot{m}$ (Mahadevan 1997).
Large values of $x_M$, as is the case in ADAFs,
imply that most of the observed synchrotron radiation originates in
the exponential tail of the single particle emissivity.
Since the electrons responsible for most of the synchrotron emission are 
relativistic (eq. [\ref{gammamaxeq}]), we can evaluate $C(\gamma)$ by using the  
synchrotron formula for $j(\nu, \gamma)$ (Rybicki \& Lightman 1979)
$$j(\nu, \gamma) =   S_0 \, \sin\alpha_p \, F\left(\nu \over \nu_c\right), $$
\beq
S_0 = {3^{1/2} (2\pi) e^2 \, \nu_B \over c}, \ \ 
\ \ \nu_c = {3 \over 2} \gamma^2 \nu_B \sin \alpha_p, \ \ \  \nu_B = {e B \over 2 \pi 
m_e c},
\label{synchem}
\eeq
where $\alpha_p$ is the angle the electron makes with the magnetic
field, $F(x)$ is an integral over modified Bessel functions (Rybicki
\& Lightman 1979, eq. [6.31c]), and $\nu_B = 2.8\times 10^6 B$ Hz is
the cyclotron frequency.

Second, an electron with Lorentz factor $\gamma_{\rm max}$
emits most of its radiation at its critical frequency $\nu_{c} \simeq
1.5 \, \gamma_{\rm max}^2 \nu_B$.  This radiation is, however, highly
self--absorbed and becomes optically thin only at a frequency $\nu_{t}
= 1.5 \theta_e^2 x_M \nu_B \approx 4.7 \nu_c$ (Narayan \& Yi 1995b;
Mahadevan 1997).  We can therefore set $\nu_t \gg \nu_c$ to good approximation. 

Finally, since the plasma is highly self--absorbed, the radiation
field, $I(\nu)$, is Raleigh--Jeans at a temperature $\theta_e$ up to
the frequency $\nu_t$, where the plasma becomes optically thin (see
Narayan \& Yi 1995b; Mahadevan 1997). This gives
\beq
I(\nu) =
\left\{ \begin{array}{l@{\quad  \quad}l}
        {2 \nu^2 m_e} \,  \theta_e & 0 < \nu < \nu_t, \\ 
        0  & \nu > \nu_t. \end{array} \right.
\eeq
Combining these properties, equation (\ref{synchceq}) becomes
\begin{eqnarray}
C(\gamma) &=& {3^{1/2} e^3 B \over m_e^2 c^4} \, \theta_e \, \nu_c \, 
\int_0^{\nu_t/\nu_c} F(x) \, dx,  \nonumber  \\
&\simeq& 1.2\times10^{-9} \, B^2 \, \theta_e \gamma^2, \nonumber \\
&\equiv& C_1 \, B^2 \theta_e \gamma^2,
\end{eqnarray}
where $C_1$ is a constant as defined above and we
have averaged over all pitch angles.

Equations (\ref{synchdir}) and (\ref{synchdiff}) are now easily evaluated to give
the systematic and diffusion timescales,
$$
t^{\rm s}_{\rm sys} =  {1 \over |d\ln \gamma / dt|} = {1 \over C_1 B^2\gamma} \, 
\left| 1 - 4 {\theta_e \over \gamma} \right|^{-1},  
$$
\beq
t^{\rm s}_{\rm diff} = 1\left/ {1 \over \gamma^2} \,{d(\Delta \gamma)^2 \over d
t} \right.  = {1 \over 2 C_1 \, B^2 \theta_e}. \label{synchtimes}
\eeq
Note that in the cyclotron limit, the systematic and stochastic coefficients
can be evaluated by recognizing that $j(\nu,\gamma) \propto \delta(\nu - 
\nu_B)$, which yields
\begin{eqnarray}
{d \gamma \over d t} &\simeq& 3 C_1 B^2 \theta_e, \nonumber  \\
{d(\Delta \gamma)^2 \over d t}  &\simeq& C_1 \, B^2 \, \theta_e \, \gamma^2 \beta^2.
\end{eqnarray}
In the limit $\beta \rightarrow 0$, the stochastic term is zero, and the 
electrons only heat up by absorbing radiation from the background field
(Ghisellini \& Svensson 1990).  Since we are concerned with thermalization 
of high energy electrons, we neglect the cyclotron limit in our discussion.

Setting $(t^{\rm s}_{\rm sys}, t^{\rm s}_{\rm diff}) < t_{\rm a}$, and using
equations (\ref{propeq}) and  (\ref{ta}), gives the accretion rates 
at which thermalization by self--absorption of synchrotron photons
is efficient
\begin{eqnarray}
\left( {\dot{m} \over \alpha^2}\right)^{\rm s}_{\rm sys} &\gsim&
7.6\times 10^{-5}\, {1 \over \gamma} \, 
\left| 1 - 4 {\theta_e \over \gamma} \right|^{-1} \, \left( {1 - \beta_{\rm adv} 
\over 0.5} \right)^{-1} \ r, \nonumber \\
\left( {\dot{m} \over \alpha^2}\right)^{\rm s}_{\rm diff} &\gsim&
3.8\times 10^{-5}\, {1 \over \theta_e} \, 
\, \left( {1 - \beta_{\rm adv} \over 0.5} \right)^{-1} \ r. \label{synchmdot}
\end{eqnarray}
For a plasma of temperature $\theta_e$, an electron with Lorentz
factor $\gamma$ can thermalize through synchrotron self--absorption if
the accretion rate is greater than the smaller of the two accretion
rates given in equation (\ref{synchmdot}).  The solid lines in figure
1 show this accretion rate as a function of the Lorentz factor
$\gamma$ for two values of $\theta_e$ at $r = 10$. For fixed
$\theta_e$, these curves can be scaled linearly with $r$ for other
radii (cf. eq. [\ref{synchmdot}]).  As Figure 1 indicates, synchrotron
self-absorption is significantly more efficient than Coulomb
collisions at thermalizing high energy electrons.  

Note that unlike electron thermalization, thermalization by
synchrotron self-absorption depends explicitly on the radius.  
The magnetic field is weaker at
larger radii which, for a fixed temperature,  
decreases the amount of synchrotron radiation.
This decreases the amount of self--absorption in the plasma and
so thermalization is less efficient, as is clear from
equation (\ref{synchmdot}).

The systematic term in equation (\ref{synchmdot}) is the dominant
thermalizing mechanism for electrons with $\gamma \sim \gamma_{\rm
max}$.  This term can be rewritten to give a condition on the radius
below which the plasma is thermal,
\beq
r^{\rm e}_{\rm th} \sim 1.3\times 10^4 \, \gamma \, 
\left( {\dot{m} \over \alpha^2}\right),
\label{reth}
\eeq
where we have taken $\gamma \gg \theta_e$. For a given $\gamma$ and
$\dot{m}/\alpha^2$, the electrons are thermal for all radii $r \lsim
r^{\rm e}_{\rm th}$.  To determine the 
accretion rates above which electrons with $\gamma \sim \gamma_{\rm max}$
are thermal throughout the flow, we set $\gamma \simeq 10 \theta_e$ and 
$r^{\rm e}_{\rm th}\simeq 1000$ in equation (\ref{reth}), which gives 
\beq
{\dot{m} \over \alpha^2} \gsim 8 \times 10^{-3} \theta_e^{-1} \simeq
4\times 10^{-2},
\label{gmdoteq}
\eeq
In the last
equality, we have set $\theta_e \sim 0.2$, as is valid for high
accretion rates (Mahadevan 1997).  To determine the accretion rates at
which the electrons cannot thermalize anywhere in the ADAF, we set
$r^{\rm e}_{\rm th} \sim 10$ and obtain
\beq
{\dot{m} \over \alpha^2} \lsim 10^{-4} \theta_e^{-1} \simeq  10^{-4},
\label{lmdoteq}
\eeq
where we have set $\theta_e \sim 1$, which is valid at small radii for
low accretion rates (\S 4.2).  
For these low accretion rates, the electrons responsible for
most of the synchrotron emission are unable to thermalize throughout the accretion
flow, and the assumption of thermal synchrotron radiation is no longer
valid. The form of the electron distribution function and the
resulting synchrotron spectrum at these low accretion rates are
determined in \S 4 and \S 5.

For accretion rates between these extremes, $10^{-4} \lsim \dot m/\alpha^2
\lsim 4 \times 10^{-2}$, the electrons will be thermal at small radii where 
synchrotron self--absorption is efficient, but will not necessarily be
thermal at larger radii.  The implications of this are considered in \S 6.
 
To conclude this section, we note that the above analysis breaks down
for electrons with large Lorentz factors ($\gamma \gg \gamma_{\rm
max}$).  This is because the radiation produced from these electrons
is not self--absorbed, and escapes the plasma freely.  The critical
Lorentz factor, $\gamma_{\rm crit}$, at which this occurs is estimated
by setting the electron's  critical frequency equal to the frequency at which the
plasma becomes optically thin, $\nu_t = 1.5 \gamma^2_{\rm crit}
\nu_B$.  This gives
\beq
\gamma_{\rm crit} =  \theta_e x_M^{1/2}, 
\eeq
which is always greater than $\gamma_{\rm max}$ for ADAFs since
$\gamma_{\rm crit} = 2^{-1/3} x_M^{1/6} \gamma_{\rm max} \gsim 2
\gamma_{\rm max}$.\footnote{However for accretion rates 
below $\sim 10^{-9}$ the plasma is no longer 
self--absorbed and there is no thermalization by synchrotron self--absorption  
($\gamma_{\rm crit} < \gamma_{\rm max}$ since $x_M < 1$).} 
For electrons with $\gamma > \gamma_{\rm crit}$
the stochastic Fokker-Planck coefficient (cf. eq.[\ref{synchdiff}]) is
zero, since there is no absorption, and the systematic term is due
solely to optically thin synchrotron cooling (cf. eq.[\ref{synchdir}]
with $C(\gamma) = 0$).  The electron distribution for $\gamma >
\gamma_{\rm crit}$ is therefore not necessarily thermal.  This,
however, has little effect on the synchrotron spectrum since most of
the emission comes from electrons with $\gamma \approx
\gamma_{\rm max} < \gamma_{\rm crit}$.
\section{Heating and Cooling of Electrons}
When thermalization by Coulomb collisions and synchrotron
self--absorption is inefficient, the evolution of the electron
distribution function is determined by the interplay between the
dominant heating and cooling mechanisms. These can be determined by
considering the electron energy equation
\beq
\rho T v {ds \over dR} = n v {d \epsilon \over dR} - q^c = q^{\rm e+} - q^-,
\label{eeq}
\eeq
$$ q^c \equiv kT v {dn \over dR}, $$
$$q^{\rm e+} = q^{\rm ie} + q^v, $$ where $s$ is the entropy of the
electrons per unit mass of the gas, $\epsilon$ is the internal energy of the 
electrons per unit mass,
$q^c$ is the compressive heating (or cooling) rate per unit volume,
and $q^{-}$ is the energy loss due to radiative cooling.  The total
external heating of the gas, $q^{\rm e+}$, is a sum of the heating via
Coulomb collisions with the hotter protons, $q^{\rm ie}$, and direct
viscous heating, $q^v$.  

Much of the previous work on ADAFs has neglected the compressive
heating of electrons and has assumed that the electron temperature
profile is given by solving $q^- = q^{\rm e+}$ (e.g., Narayan \& Yi,
1995b).  We show below that this is likely to be a poor approximation
except for systems near the critical accretion rate.  This point has
already been  made by Nakamura et al. (1997), who first
recognized the importance of compressive heating in determining the
electron temperature profile.

As long as $q^{\rm e+} \ll q^c$ and $q^- \ll q^c$, the right hand side
of equation (\ref{eeq}) is small compared to each term on the left
hand side.  All of the compressive heating therefore goes into changing the
internal energy of the gas, with nearly constant entropy.  We now show
that both of these conditions are satisfied at low accretion rates, 
so that the electrons in ADAFs compress adiabatically as they accrete
onto the central object. 

\subsection{Heating}
The dominant heating mechanism for the electrons is obtained by
comparing the relative magnitudes of $q^c$, $q^{\rm ie}$, and $q^v$.
Using equations (\ref{propeq}) and (\ref{eeq}), the compressive
heating rate is
\beq
q^c \simeq 5 \times 10^{17}  \dot m m^{-2} T_9 r^{-3} \ \ 
{\rm erg \ cm^{-3} \ s^{-1}}.
\eeq
The energy transfer rate between
protons and electrons via Coulomb collisions is (Stepney \& Gilbert 1983)
\begin{eqnarray}
q^{\rm ie} &\simeq& 2.05 \times 10^{-17} n_e n_i (T_i - T_e) T^{-3/2}_e \nonumber \\
&=& 2.5 \times 10^{21} \alpha^{-2} {\dot m}^2 m^{-2} T_9^{-3/2} r^{-4}  \ \ 
{\rm erg \ cm^{-3} \ s^{-1}},
\end{eqnarray}
where we have assumed that $T_e \leq 10^{10}$K so that a
non-relativistic formula is reasonably accurate.  The last equality
follows from using equation (\ref{propeq}), and setting $T_e \ll T_i$,
$n_e = n_i$.  

The viscous heating of the electrons is determined by
allowing for a fraction $\delta \sim m_e/m_p \approx 1/2000$ of the
viscously generated energy to be transferred directly to the electrons,
so that (Mahadevan 1997)
\beq
q^v \simeq 3 \times 10^{17} {\dot m} m^{-2} r^{-4} \delta_o \ \ 
{\rm erg \ cm^{-3} \ s^{-1}},
\eeq
where $\delta_o = 2000 \delta$.  
Both compressive and viscous heating are proportional to the
accretion rate and their ratio is given by
\beq
q^c/q^v \sim 2 \, \delta_o^{-1} \, T_9 \, r,
\eeq
which is independent of the global parameters of the ADAF. In the inner 
regions of the accretion flow ($r \sim 3$), $T_9 \sim 5$, while for larger 
radii the  
temperature decreases roughly as $r^{-2/3}$ (\S 4.2).
Compressive heating of the electrons is therefore always more important
than direct viscous heating provided that $\delta$ is not substantially
larger than $10^{-3}$.

Coulomb heating of the electrons by the protons is, however, a two
body process and becomes less important for small $\dot m$.  From
the expressions for $q^c$ and $q^{\rm ie}$ given above, it follows that
compressive heating dominates over Coulomb heating provided that
\beq
{\dot m } \lsim 2 \times 10^{-4} \alpha^2 T^{5/2}_9 r.
\label{mdotc}
\eeq
Comparing this with equations (\ref{eemdot}) and (\ref{synchmdot})
shows that for low $\dot{m}$, where both Coulomb collisions and
synchrotron self--absorption are unable to thermalize the plasma,
compressive heating is the dominant heating mechanism.
\subsection{Cooling}
 
Synchrotron cooling can be an important energy loss mechanism in ADAFs
because of the assumption of near equipartition magnetic fields.  In
fact, for the low mass accretion rates of interest here, synchrotron
radiation is the dominant source of cooling (Narayan
\& Yi, 1995b; Mahadevan 1997).  In \S 2 we showed that the 
thermalization time through synchrotron self--absorption is of order
the electron cooling time and that self-absorption is inefficient at
thermalizing the electrons for $\dot m \lsim 10^{-4}
\alpha^2$ (cf. Figure 1).  For these low accretion rates electrons therefore
cannot radiate a significant fraction of their energy in an accretion
time.  Since the dominant source of the electron's energy is
compressive heating ($q^c$) this implies that $q^- \ll q^c$.

This argument relies on the accuracy of our estimate of the electron
temperature in the accretion flow, since higher temperatures would
lead to more efficient synchrotron cooling.
Since previous work has neglected compressive heating, however, we
must verify that electron temperatures of $\sim 10^9$ K are still
valid when the electron temperature profile is determined by adiabatic
compression.

For adiabatic compression, the temperature profile is given by $T
\propto \rho^{\Gamma - 1}$, where $\Gamma$ is the ratio of the gas
specific heats and $\rho \propto r^{-3/2}$.  Since the internal energy
of the electron gas includes a contribution from the magnetic field
energy, $\Gamma = (8 - 3\beta_{\rm adv})/(6 - 3\beta_{\rm adv})$ for
non--relativistic electrons (Esin 1997).  For an equipartition
magnetic field, $\Gamma = 13/9$, which gives a temperature profile of
$T \propto r^{-2/3}$.  $\Gamma = 4/3$ is still valid in the
relativistic limit, in which case $T
\propto r^{-0.5}$.  For $T_e \approx T_p$ at $r \approx 10^4$, 
the above scalings imply that the electron temperature near the last
stable orbit is $\approx 5 \times 10^9$ K, which is comparable with previous
estimates.
\subsection{The Distribution Function of an Adiabatically Compressed Gas}
For $\dot{m}\lsim 10^{-4} \alpha^2$, Coulomb collisions and
synchrotron self--absorption are unable to thermalize the electrons.
The electrons therefore do not interact with one another and can be
well approximated as a collisionless gas.  Furthermore, at these low
accretion rates, the electrons compress nearly adiabatically as they
accrete onto the central object.

An adiabatic change is one in which the properties of a system vary
slowly compared to the characteristic timescales of the system.
During such a change, the system's phase space volume is a conserved
quantity (Landau \& Lifshitz, {\em Mechanics} \S 49). We use this
property to determine the evolution of the distribution function of a
collisionless electron gas subjected to adiabatic changes in its
volume.  We emphasize that the discussion which follows is not specific
to an ADAF, but applies to any collisionless plasma which undergoes
adiabatic changes.

The adiabatic compression of a collisionless gas can be viewed as the
independent adiabatic compression of each particle in the gas, since
there is no coupling between the particles.  By adiabatic invariance,
the quantity $p^3 \rho^{-1}$ is conserved for each particle in a
monatomic gas, where $p$ is the momentum of the particle and $\rho$ is
the density of the gas.\footnote{$p^3
\rho^{-1}$ is also the adiabatic invariant for particles moving in an
isotropically tangled, flux frozen, magnetic field.}  The evolution of
the distribution function is therefore given by the transformation $p
\rightarrow a p$, where $a = (\rho_f/\rho_i)^{1/3}$, and $\rho_{i(f)}$
denotes the initial (final) gas density.  We emphasize that this
transformation is valid for all momenta $p$, both non--relativistic and relativistic.

To determine how the distribution function of a non-relativistic
thermal gas of temperature $\theta_e$ changes under an adiabatic
compression, consider the Maxwell--Boltzmann distribution function
\beq
n(p)dp = {4 \pi p^2 \over (2 \pi m^2 c^2 \theta_e)^{3/2}} \exp\left(-{p^2 \over 2
m^2 c^2 \, \theta_e}\right)dp.
\label{nrmb}
\eeq
For now, assume that the compression factor, $a$, is such that the
particle momenta after compression remain non-relativistic.
Substitution of the transformation $p \rightarrow a p$ into equation
(\ref{nrmb}) shows that the form of the distribution function remains
the same. The gas is still thermal, but is now characterized by a
temperature $\theta_e^{\prime} = a^2 \theta_e = (\rho_f/\rho_i)^{2/3}
\theta_e$, which is the usual result for the adiabatic compression of
a $\Gamma = 5/3$ gas.  A thermal distribution is therefore maintained
even though the gas is collisionless.

Similarly, for a relativistic thermal gas
characterized by the Maxwell--Boltzmann distribution function,
\beq
n(p)dp = {p^2 \over m^3 c^3 \theta_e K_2(1/\theta_e)} \exp\left(-{p \over m c 
\theta_e}\right) dp \label{rmb}
\eeq
setting $p \rightarrow a p$ and assuming that the final momenta are
relativistic shows that the gas remains thermal with a final
temperature $\theta_e^{\prime} = a \theta_e = (\rho_f/\rho_i)^{1/3}
\theta_e$.  This is the usual result for the adiabatic compression of a
$\Gamma = 4/3$ gas.

Under adiabatic compression, an initially non--relativistic
(relativistic) thermal gas therefore remains a non--relativistic
(relativistic) thermal gas, as long as the post compression momenta
remain non--relativistic (relativistic).  Consider, however, a
non--relativistic thermal gas (eq. [\ref{nrmb}]), which is compressed
by a sufficient amount that the final momenta of the particles are now
relativistic.  Since the transformation $p
\rightarrow a p$ is valid regardless of the magnitude of the momenta,
the post compression distribution function is given by
\beq
n(p)dp = {4 \pi p^2 \over (2 \pi m^2 c^2  a^2 \theta_e)^{3/2}} \exp
\left(-{p^2 \over 2
m^2 c^2  \, a^2 \theta_e}\right)dp.
\label{cnrmb}
\eeq
This has the form of a non-relativistic Maxwellian, even though the
particle energies can be highly relativistic.  Therefore a
collisionless gas which is adiabatically compressed across the
non--relativistic/relativistic boundary does not maintain a thermal
distribution.\footnote{Note that for photons, which are massless,
there is no such boundary, and so adiabatic changes always maintain a
thermal distribution, a result which is, of course, well known in
cosmology.}  In particular, the relativistic high energy tail of the
adiabatically compressed distribution function falls off as a Gaussian
rather than as an exponential (cf. eqs. [\ref{cnrmb}], [\ref{rmb}]).

In order to assess the implications of this new distribution function,
we compare it with a thermal distribution function which has the
same average energy (the ``best fit'' thermal distribution).
The average energy of an adiabatically compressed gas in the
relativistic limit is (in units of $m_e c^2$)
\begin{eqnarray}
\langle \gamma \rangle &=&  {4 \pi \over (2 \pi \, a^2  \, \theta_e)^{3
/2} }
\int_0^{\infty}\gamma^3 \exp \left( - {\gamma^2 \over 2 \, a^2 \, \theta_e} \right)
\, d \gamma, \nonumber \\
&=& {4 \over (2 \pi )^{1/2} } \, a \, \theta_e^{1/2}.
\label{mg}
\end{eqnarray}
The average energy of the gas increases as $a$ in the relativistic
regime, rather than as $a^2$ as in the non--relativistic regime.
Since the average energy of a thermal relativistic gas is $\langle
\gamma \rangle = 3
\theta_{MB}$, the ``best fit'' thermal gas has
\beq
\theta_{MB} = \left( {8 \over 9 \pi} \right)^{1/2} \, a \, \theta_e^{1/2}. \label{rtoace}
\eeq

Figure 2a shows the distribution function for a gas which is
adiabatically compressed by a factor of $a = 20$, from an initially
thermal distribution with $\theta_e = 0.01$.  Superimposed in the
figure is the distribution function for a thermal gas which has the
same mean energy as the adiabatically compressed gas ($\theta_{MB} \simeq
1 $).  As expected, the high energy tail of the adiabatically
compressed gas is strongly suppressed with respect to the thermal gas.

A relativistic collisionless gas that adiabatically expands across the
relativistic/non--relativistic boundary will also not maintain a
thermal distribution.  Fig 2b shows the distribution function for a
gas which adiabatically expands by a factor of $a = 0.01$ from an
initially thermal distribution with $\theta_e = 10$.  Superimposed in
the figure is the distribution function for a thermal gas which has
the same mean energy ($\theta_{MB} \simeq 0.04$).  The high energy tail
of the adiabatically expanded gas is overpopulated with respect to the
thermal gas since the distribution function falls of as an
exponential, rather than as a Gaussian, in the non--relativistic
limit.

Finally, we note that it is not necessarily valid to assume that the electrons have
an adiabatically compressed distribution function
just because thermalization is inefficient.  This is because the 
timescale for thermalization as determined 
in \S 3 is not the same as the timescale on which an initially 
non--thermal distribution function changes substantially. 
As we discuss in
Appendix C,  it is only strictly true that the electrons have the adiabatically 
compressed distribution function out to $\gamma \sim \gamma_{\rm max}$ if 
the accretion rate is  $\sim 10$ times
smaller than the thermalization accretion rate given in \S 3 (cf. eqs. [\ref{synchmdot}]). 
However, to accurately determine the form of the distribution function 
in this regime requires solving the complete Fokker--Planck equation, which is
beyond the scope of this paper.  In what follows, we assume that, 
for this narrow range in accretion rates, the electrons are 
well approximated as having an adiabatically compressed distribution function. 
\section{The Synchrotron Spectrum from  an Adiabatically Compressed Gas}
For an adiabatically compressed gas the large decrease in the number
of particles at high energies has important consequences for the
synchrotron emission from ADAFs.  This is because most of the
synchrotron emission originates from high energy electrons in the tail
of the Maxwell--Boltzmann distribution.  A lack of these high energy
electrons will substantially reduce the synchrotron luminosity.

To determine the synchrotron emission from this new distribution,
we first calculate the synchrotron source function, $S_{\nu}$,
and then calculate the emergent intensity.  In the present analysis,
we assume that all of the electrons are relativistic so that the
distribution function (eq. [\ref{cnrmb}]) takes the form
\beq
n(\gamma) \, d\gamma = {4 \pi \over (2 \pi \, a^2 \theta_e)^{3/2}} \, \gamma^2
\exp\left(- {\gamma^2 \over 2 \, a^2\theta_e}\right) \, d\gamma.
\eeq
Using equation (\ref{synchem}), the emissivity, $\epsilon(\nu)_{\rm
AC}$, and the absorption coefficient, $\alpha(\nu)_{\rm AC}$, are
given by
\begin{eqnarray}
\epsilon(\nu)_{\rm AC} \, d\nu &=& \int j(\nu,\gamma) \, n(\gamma) \, d\gamma,
                \nonumber \\
&=& E_0 \,  {\chi \over a^2 \theta_e} \,
I^{\prime}\left(\xmp , 2\right) \, d\nu \ \ \ \mbox{ergs s$^{-1}$ Hz$^{-1}$},
        \label{epsilonAC}\\
\alpha(\nu)_{\rm AC} \, d\nu &=&
- {1 \over 8 \pi \,   m_e \nu^2} \int \gamma p \, j(\nu, \gamma)
        \, {\partial \over \partial\gamma} \left[ {n(\gamma)  \over \gamma p }
\right]  \, d\gamma,  \nonumber \\
&=& E_1 \, {\chi^{-1} \over (a^2 \theta_e)^{3/2} } \,
I^{\prime}\left(\xmp, 3\right) \, d\nu \ \ \ \mbox{cm$^{-1}$},
        \label{alphaAC}
\end{eqnarray}
where $p \rightarrow m_e\gamma c$ in the last equality.  We have
defined
$$I^{\prime}(x, n) = {1 \over 4 \pi } \int I\left({x \over \sin\alpha_p}, n\right) \,
d\Omega_p,$$ 
$$I(x, n) \equiv {1 \over x} \, \int_0^{\infty} \,
F\left[ {x \over z^2} \right] \, z^n \, \exp(-z^2) \, d\gamma, $$ and
\beq
E_0 \equiv {8\pi^{1/2} e^2 \nu_B \over \sqrt{3} c}, \ \ \
E_1 \equiv {2^{1/2}\, e^2 \over  \sqrt{3\pi } \,  c m_e \nu_B}, \ \ \
\xmp \equiv { \chi \over 3 \, a^2 \theta_e}, \ \ \ \chi \equiv {\nu \over \nu_B},
\label{ioxm}
\eeq
to maintain a similarity between the equations presented here and
those for thermal synchrotron emission (Pacholczyk 1970; see also
Mahadevan et al. 1996).  $I^{\prime}(\xmp, n)$ is $I(\xmp, n)$
integrated over all particle directions.  The asymptotic expansions
for $I(\xmp, n)$ and $I^{\prime}(\xmp, n)$ at large and small $\xmp$
are given in Appendix A.

The source function is determined using the limiting expressions for
$I^{\prime}(x, n)$ (Appendix A), which gives
\beq
S_{\nu} =
{\epsilon(\nu)_{\rm AC} \over
4\pi \, \alpha(\nu)_{\rm AC}} =
\left\{ \begin{array}{l@{\quad  \quad}l}
        {2^{1/2} m_e \nu^2} \,  (a^2 \theta_e)^{1/2} & \xmp \ll 1, \\
        2^{1/2} 3^{1/4} m_e \nu_B^{1/4} \, \nu^{7/4} \, (a^2 \theta_e)^{3/4} & \xmp \gg 1.
\end{array} \right.
\eeq
The source function for a thermal gas in the Raleigh--Jeans limit is
$S_{\nu} = 2 m_e \nu^2 \, \theta_e$.  For low frequencies, $\xmp \ll
1$, the source function for the adiabatically compressed gas has the
same frequency dependence as that of a thermal gas, but the normalization
is larger by a factor of $3 \sqrt{\pi}/4 \simeq 1.33$ because there is
an excess of low energy electrons (see Figure 2a).  For large
frequencies, $\xmp \gg 1$, however, the frequency dependence of the
source function for the adiabatically compressed gas is different than
that of a thermal gas, and varies as $\nu^{7/4}$.

The emissivity for an adiabatically compressed gas at large
frequencies is given by
\beq
\epsilon(\nu)_{\rm AC} \, d\nu = 1.6\times 10^{-28} \, \nu_B  \, 
\xmp^{3/4} \, \exp\left(-2 \xmp^{1/2} \right) \, d\nu, \ \ \ \mbox{ergs s$^{-1}$ Hz$^{-1}$,}
\eeq
where we have used the expression for $I^{\prime}(\xmp, n)$ for $\xmp
\gg 1$ (Appendix A).  For a thermal gas, the corresponding emissivity
is (Mahadevan et al. 1996, eqs. [11], [A10]),
\beq
\epsilon(\nu)_{\rm Th} \, d\nu = {3.4 \times 10^{-28}}\, 
{\theta_e^2 \nu_B\, \over K_2(1/\theta_e)} \, x_M^{5/6} \, \exp\left( -1.8899 x_M^{1/3} 
\right) \, d\nu, \ \ \  \mbox{ergs s$^{-1}$ Hz$^{-1}$,}
\eeq
where $$x_M \equiv {2 \chi \over 3 \theta_e^2 }.$$ The primary
difference in the emissivity is the frequency dependence in the
exponential.  The exponential in the thermal spectrum falls off as
$\nu^{1/3}$, while it falls more rapidly, as $\nu^{1/2}$, for an
adiabatically compressed gas. This is expected because there is a lack
of high energy particles in the adiabatically compressed gas as compared to a
thermal gas with the same average energy.

Figure 3a compares the total synchrotron luminosity from a sphere of
radius $R$ for a thermal and an adiabatically compressed electron gas
with the same average energy ($\theta_e = 0.5$ for the thermal gas).
We set $R \simeq 7 \times 10^{11}$ cm, $B \simeq 200 $ Gauss, and $n_e
\simeq 10^{10} $ cm$^{-3}$, which are the parameters at ten Schwarzschild 
radii for a $2\times 10^{6}$ solar mass black hole accreting at $\dot
m \simeq 10^{-5}$.  The solid line shows the spectrum for the adiabatically
compressed gas, the dashed line shows the spectrum for the thermal
gas, and the dotted line shows the optically thin emission that would
result in each case were the plasma not self-absorbed. The
adiabatically compressed gas has a significantly smaller synchrotron
luminosity and becomes optically thin at a much lower frequency than
the thermal gas.  The adiabatically compressed gas is, however, still
highly self-absorbed, thus validating our assumption that most of the
emission comes from the exponential tail of the single particle 
emissivity.

Since the electron number density and magnetic field strength vary with
radius, the total synchrotron spectrum from an ADAF is
obtained by summing the individual spectrum from each radius in the
accretion flow (e.g., Narayan \& Yi, 1995b).  An example of this is
shown in Figure 3b for an adiabatically compressed gas (solid line) and a 
thermal gas (dashed line), taking $m = 2\times 10^6$,
$\dot{m} = 10^{-5}$, and using equation (\ref{propeq}) for the number
density and magnetic field as a function of radius.  In this calculation, we have
assumed for simplicity, and for comparison with previous work, that
the internal energy of the electrons is constant throughout the
accretion flow ($\theta_e = 0.5$ for the thermal gas).  As Fig. 3b shows,
the total luminosity from the adiabatically compressed gas is much
lower than that of the thermal gas since there are fewer high energy
electrons.  The slope of the spectrum is, however, the same for the
thermal and adiabatically compressed gases because both are highly
self--absorbed and have very similar source functions.

The assumption of constant electron internal energy is valid at high
$\dot m$ where the cooling is efficient, as has been demonstrated by a
number of previous calculations (e.g. Narayan \& Yi, 1995b). For the
lower accretion rates of interest here, where adiabatic compression
determines the electron temperature profile, a constant electron
temperature is an invalid assumption.  Future calculations of
synchrotron emission from low $\dot{m}$ systems must self consistently
calculate the evolution of the electron internal energy during
adiabatic compression.  The details of this calculation are given in
Appendix B.

Finally, we note that the slope  
of the synchrotron spectrum in high $\dot{m}$  ADAFs is very similar to the
characteristic radio slope observed in many black hole candidates (Mahadevan 1997). 
This is, however, a direct consequence of the assumed constancy of $\theta_e$, which 
is not in general valid.
Therefore we expect that low $\dot m$ ADAFs will have (1) an unexpectedly
low radio luminosity because of the absence of high energy electrons
in the distribution function and (2) a radio slope that differs from
that of high $\dot m$ systems because the assumption of constant
electron internal energy is no longer valid.
\section{Discussion \& Conclusions}
In an advection dominated accretion flow, the timescale for protons
and electrons to exchange energy by Coulomb collisions is sufficiently
long compared to the accretion time that the two species are essentially
decoupled.  The protons and electrons can therefore have different
temperatures and, if non--thermal, different distribution functions.
The precise form of the electron and proton distribution functions is
crucial for comparing calculated spectra from ADAFs with observations.

Coulomb collisions among the protons are too inefficient to force the
protons to be thermal, since the timescale for thermalization is much
longer than the accretion time (\S2).  The proton distribution
function is therefore determined by the viscous heating mechanism,
which is unknown.  Since the protons are marginally relativistic,
Mahadevan et al. (1997) have shown that an ADAF produces a substantial
$\gamma$--ray flux, which is created from the decay of neutral pions
produced through proton--proton collisions.  In particular, they have
shown that the luminosity and shape of the gamma-ray spectrum differs
dramatically if the protons have a thermal or a power--law
distribution.  Comparison of the predicted $\gamma$--ray spectrum with
observations can therefore determine the form of the proton
distribution function and, in principle, provide a means of
understanding the viscous heating mechanism in hot accretion flows.
Mahadevan et al. (1997) have shown that the EGRET observations of the
source 2EG J1746--2852 (Merck et al. 1996), which is coincident with
the Galactic Center Sgr A$^*$, are in good agreement with the
predicted $\gamma$--ray spectrum from an ADAF, provided that the
protons have a power--law distribution of the form $E^{-2.75}$.  This
is similar to the cosmic--ray proton energy distribution, which leads
to the speculation that the heating mechanism responsible for
accelerating cosmic rays might also be at work in ADAFs.

We have considered thermalization of electrons through both Coulomb
collisions and self--absorbed synchrotron radiation.  The latter
process is significantly more important, since it can operate even
when the electrons are effectively collisionless.  We find that, for
high accretion rates, $\dot{m} \gsim 4 \times 10^{-2} \alpha^2$, the
electrons can efficiently exchange energy throughout the ADAF by 
both Coulomb collisions and the
absorption and emission of synchrotron photons (\S 3).  To conclude
from this that the electrons are thermal also requires that thermalization
proceed on a shorter timescale than the heating/cooling of the
electrons.  This ensures that the electron distribution function can
relax to a Maxwellian more rapidly than the heating/cooling induces
modifications in the distribution function.  Since synchrotron radiation
is the  dominant cooling mechanism, and also leads to thermalization because it is 
highly self--absorbed, this criterion is trivially satisfied
for the cooling of the electrons in ADAFs.

Modifications of the electron distribution function by heating are
perhaps a more viable concern.  For the high mass accretion rates
where thermalization appears to be efficient ($\dot{m}
\gsim 4 \times 10^{-2} \alpha^2$), the electrons are primarily heated by
Coulomb collisions with the hotter protons, provided that most of the
viscously generated energy is transferred to the protons (\S 4.2).
Since electron heating by the protons occurs on a timescale long
compared to the thermalization timescale by electron-electron Coulomb
collisions (\S 2.2), which is in turn long compared to the thermalization
timescale by synchrotron self-absorption (\S 3.2), we conclude that heating of
the electrons by the protons will not significantly modify the
electrons from a thermal distribution.\footnote{Nonthermal heating of
the electrons by wave-particle interactions, reconnection, etc. could,
perhaps, modify the electrons from a thermal distribution. We have,
however, as mentioned in the introduction, 
restricted our analysis to those processes whose role in
ADAFs is relatively well understood, and so have not considered these effects.  
In addition, we note that the
applicability of some of these non-thermal heating mechanisms to the
nearly collisionless plasma in ADAFs is unclear.}  Therefore, for $\dot m
\gsim 4
\times 10^{-2}
\alpha^2$, where self-absorbed synchrotron radiation allows for
efficient energy exchange among the electrons throughout the accretion
flow, the electrons are likely to be thermal.

We find that for lower accretion rates, $\dot{m} \lsim 10^{-4}
\alpha^2$, thermalization is inefficient throughout the ADAF and thus the 
electrons are not necessarily thermal.  In this regime, the heating
and cooling mechanisms determine the electron distribution function.
At these low accretion rates, cooling is very inefficient and
compressive heating of the electrons is the dominant heating mechanism
(\S 4).  The evolution of the electron distribution function is therefore
determined by adiabatic compression.  Using the
principle of adiabatic invariance,  we find that the electron distribution
function evolves by the transformation $p
\rightarrow a p$, where $p$ is the electron momenta (relativistic 
{\em or} non-relativistic) and $a$ is a function of the electron
density.  For ADAFs, in which the internal energy of the gas contains
a contribution from the magnetic field energy, the expression for the
compression factor, $a$, is more complicated than that given in \S
4.3.  The precise form of $a(\rho)$ does not, however, effect any of the
conclusions of this paper.  Nonetheless, in  Appendix B we derive the form 
of $a(\rho)$ under the assumption that the magnetic field always contributes a constant
fraction of the total pressure.

The above transformation for the electron momenta under adiabatic
compression yields the distribution function given in equation
(\ref{cnrmb}), which is non--thermal if the electrons
have relativistic energies. In particular, the high energy tail of an
adiabatically compressed (expanded) gas is strongly suppressed (enhanced) 
with respect to a thermal gas of the same average energy (cf. Fig. 2).  
We emphasize that this is not specific to an ADAF.  Any collisionless plasma
which undergoes an adiabatic change will have a non--thermal 
distribution function if the particles' momenta cross the relativistic/
non--relativistic boundary.

Our derivation of the distribution function of an adiabatically
compressed electron gas is based on the assumption that the electrons
are thermal at large radii.  In many models where ADAFs have been
successfully applied, the ADAF forms from a thin accretion disk at $r
\sim 10^4$ (Narayan, McClintock, \& Yi 1996).  In this case, the
electron distribution function is Maxwellian at large radii by virtue
of the low temperature and large density in the thin accretion disk (cf. eq. [\ref{mdth}]).
In the event that an accretion disk is not present, as may be the case
for super massive black holes at the centers of elliptical galaxies
which accrete the surrounding gas through Bondi accretion, the gas at
large radii is probably still thermal.  This follows
from the X--ray gas profiles of the centers of elliptical galaxies, 
 which indicates that the gas is 
thermal with a temperature $\sim 10^{7.5}$ K (eg. Trinchieri et al. 1988).  

For accretion rates such that $10^{-4} \lsim (\dot{m}/\alpha^2) \lsim
4 \times 10^{-2}$ the electrons are thermal for only part of the
accretion flow.  The initially thermal electron gas is adiabatically
compressed down to a radius $\sim r^{\rm e}_{\rm th}$
(cf. eq. [\ref{reth}]), with no means of thermalizing by synchrotron
self-absorption or Coulomb collisions.  
For $r \lsim r^{\rm e}_{\rm th}$, however, electron
thermalization by synchrotron self-absorption becomes efficient
because the magnetic field is stronger and there is more synchrotron
radiation in the plasma. 

As the accreting gas passes through the ``transition'' radius, $r^{\rm
e}_{\rm th}$, the electrons switch from an adiabatically compressed
distribution function to a thermal one.  This may substantially modify
the predicted spectra from some ADAFs.  For emission
mechanisms in which the bulk of the emission comes from electrons with
roughly the mean energy of the plasma, such as bremsstrahlung, the
modified distribution function is unlikely to have a pronounced effect
on the spectrum.  Synchrotron radiation, on the other hand, is highly
self--absorbed, with most of the emission coming from electrons in the
high energy tail of the distribution function.  Therefore, even when the
mean electron energy is non--relativistic, the non--thermal distribution
function due to adiabatic compression modifies the synchrotron
spectrum substantially.  
In \S5 we explicitly calculated the synchrotron spectrum
from an adiabatically compressed electron gas. 
The  synchrotron  
luminosity can be significantly smaller than that of a thermal
electron gas with the same energy because there is a deficit of high
energy particles in the tail of the distribution function (cf. Fig.
3). We note that, while the applications in this paper have been to 
ADAFs, the synchrotron emissivity and source function 
  calculated  in \S5 apply  to 
any adiabatically compressed gas.   

Previous work in calculating the spectra from ADAFs has assumed that
the electrons are thermal at all radii.  These systems include 
A0620, V404Cyg (Narayan et al. 1996), and NGC 4258 (Lasota et
al. 1996), where the accretion rates all satisfy  $\dot m \gsim
10^{-2} \alpha^2$.  At these high accretion rates the
assumption of thermal electrons is reasonably valid and the predicted
spectra from these systems is unchanged by the present work.  For
systems with lower accretion rates, however, the change in the
distribution function described here might considerably modify the
predicted spectrum.

In particular, the ADAF model of Sgr A$^*$ (Narayan et al. 1995) should be reexamined 
since the estimated accretion rate falls below $10^{-2} \alpha^2$.  Narayan
et al. (1995) obtained good agreement with the observed spectrum of Sgr A$^*$ 
 using $m = 7\times 10^5$, $\dot{m} \simeq 3 \times 10^{-3}\alpha^2$, and an
equilibrium temperature of $\theta_e \sim 1.5$.   Using these parameters and
equation  (\ref{reth}) gives $r^{\rm e}_{\rm th} \simeq 600$. For $r \lsim 600$
the assumption of thermal electrons is no longer valid.
Since each frequency in the radio spectrum corresponds to emission from a
particular radius, $\nu \propto r^{-5/4}$ (Mahadevan 1997),
and the maximum synchrotron frequency  for Sgr A$^*$ is $\sim 10^{12}$ Hz 
(Narayan et al. 1995), the emission from $r \gsim 600$
corresponds to frequencies $ \lsim 10^9$ Hz.
For these frequencies the model of Narayan et al. (1995) may be inconsistent, but 
since almost all of the data is for $\nu \gsim 10^9$ Hz
(see Fig. 1  in Narayan et al. 1995), the possibility of non--thermal
electrons is unlikely to be important for this system. 

A more promising application of the present work is in modeling the $3
\times 10^7 \ M_{\odot}$ black hole in M31 (Kormendy
\& Richstone 1995).  The accretion rate in this system is estimated to
be $\dot{m} \sim 10^{-4}$ (Goodman \& Lee 1989),
which is low enough that thermalization at all radii is  unlikely.
Using an ADAF model with $\alpha = 0.3$ and $T_e \sim T_p
(r/10^4)^{1/3}$ (\S 4.2) gives $r^{\rm e}_{\rm th} \sim 100$.  For $r
> 100$ the synchrotron spectrum would need to be calculated using the
adiabatically compressed distribution function and would differ
considerably from the spectrum calculated from a thermal distribution
of electrons. Since lower frequencies in the synchrotron spectrum
corresponds to emission from larger radii, 
the emission at frequencies $ \sim 10^{10}$ Hz may 
be greatly reduced because of the new electron distribution function
(cf. Fig. 3b).  In particular, this may explain the extremely low radio
flux of $\sim 10^{32} $ ergs s$^{-1}$ at $\nu \sim 10^{10}$ Hz (Crane
et al. 1992) in M31, but detailed numerical
calculations are required to confirm this hypothesis.

Another potential application of this work is in determining whether
elliptical galaxies host dead quasars (Fabian \& Canizares 1988).  As
suggested by Fabian \& Rees (1995), ADAFs might allow elliptical
galaxies to host $\gsim 10^8 \ M_{\odot}$ black holes and still remain
dim.  By varying the mass of the central black hole, Mahadevan (1997;
see also Reynolds et al. 1996) has given spectra, which agree with the
radio and X--ray upper limits, for a few of the nearby elliptical
galaxies. He found that elliptical galaxies can host black holes up to
$\lsim 10^{9.5} \ M_{\odot}$, provided that they accrete via ADAFs.
Interestingly, the constraint on the mass does not come from the X--ray
data, but rather from the radio flux upper limit.  Since these
galaxies have estimated accretion rates $(\dot{m}/\alpha^2) \lsim 4
\times 10^{-4} \, M_8$ (Mahadevan 1997, eqn.[56]), where $M_8 = 10^8 \
M_{\odot}$, the radio spectrum may be reduced due to the adiabatically
compressed electron distribution function.  This would lead to a further increase
in the allowed mass of the central black holes, therefore completely 
eliminating the dead quasar host problem highlighted by Fabian \& 
Canizares (1988).

A final application of this work is in determining accurate 
emission spectra from isolated black holes
in the disk and halo of our Galaxy, which accrete from the interstellar medium.
Ipser \& Price (1982; 1977) have calculated spectra from these systems under the 
assumption  that the black holes accrete by Bondi accretion and that the 
spectra is dominated by optically thin thermal synchrotron radiation.
Even a small amount of angular 
momentum in the accreting gas precludes Bondi accretion, 
and so the gas is more likely to accrete onto the black hole as an ADAF.  
Furthermore, since these systems are likely to have low accretion rates
($\dot{m} \lsim 10^{-5}$), the electrons  are unable to thermalize and their distribution 
function is probably determined by adiabatic compression.  The spectra
from these systems will therefore be considerably different,  with lower fluxes and 
luminosities  peaking in different wave bands,  than those obtained using 
a thermal distribution of electrons.   
Since the spectra from ADAFs are robust and well understood, 
detailed calculations of the spectra from isolated black holes, taking
into account the non--thermal electron distribution function, would
be quite valuable for the potential detection of these systems by the
Sloan Digital Sky Survey.  In addition, if no candidates are found, the 
ADAF spectrum will considerably modify the limits on halo black hole populations set
by Heckler \& Kolb (1996).

\noindent{\it Acknowledgments.}  We thank Ramesh Narayan for many 
insightful and stimulating discussions throughout this work,  and for 
comments on the manuscript.  We also thank Ann Esin, Charles Gammie, 
Zolt\'an Haiman, Jeffrey McClintock, and George Rybicki for useful discussions.  
RM was supported by NSF grant AST 9423209 and EQ was supported by an NSF Graduate 
Research Fellowship.
\newpage
\begin{appendix}
\section{Asymptotic Formulae for $I(x,n)$ and $I^{\prime}(x,n)$}
\subsection{$I(x,n)$}
The definition of $I(x,n)$ is given in equation (\ref{ioxm}):
\beq
I(x, n) \equiv {1 \over x} \, \int_0^{\infty} \,
F\left[ {x \over z^2} \right] \, z^n \, \exp(-z^2) \, d\gamma.
\eeq
This can be evaluated using the limiting forms for 
$F(x)$  (Rybicki \& Lightman 1979; eqns. [6.34a,b]), which gives
\begin{eqnarray}
I(x, 2) &\rightarrow&  {2^{2/3} \pi  \over \sqrt{3}} \, 
{\Gamma(7/6) \over \Gamma(1/3)} \, x^{-2/3} = 
1.01\, x^{-2/3}, \ \ \ x \ll 1, \nonumber \\
I(x, 3) &\rightarrow&  {2^{2/3} \pi  \over \sqrt{3}} \,
{\Gamma(5/3) \over \Gamma(1/3)} \, x^{-2/3} = 0.98 \, x^{-2/3}, \ \ \ x \ll 1, 
\end{eqnarray}
and
\begin{eqnarray}
I(x, 2) &\rightarrow& {\pi \over 2^{3/2}} \, x^{-1/4} \, 
	\exp\left( - 2 x^{1/2} \right) =  
1.11 \, x^{-1/4} \, \exp\left( - 2 x^{1/2} \right), \ \ \ x \gg 1, \nonumber \\
I(x, 3) &\rightarrow& {\pi \over 2^{3/2}} \, \exp\left( - 2 x^{1/2} \right) = 
1.11 \, \exp\left( - 2 x^{1/2} \right), \ \ \ x \gg 1.
\end{eqnarray}
\subsection{ $I^{\prime} (x,n)$}
The function $I^{\prime}(x,n)$ is the angle averaged value of $I(x,n)$ and is defined by 
\beq
I^{\prime}(x, n) = {1 \over 4 \pi } \int I\left({x \over \sin\alpha_p}, n\right) \, 
d\Omega_p.
\eeq
For $x \ll 1$ the integration is the same as done by Mahadevan et
al. (1996), which gives
\begin{eqnarray}
I^{\prime}(x, 2) &\rightarrow&
 {2^{-1/3} \pi  \over \sqrt{3}} \,
{\Gamma(7/6) \, \Gamma(1/2) \over \Gamma(11/6)} \, x^{-2/3} =
0.85 \, x^{-2/3}, \ \ \ x \ll 1, \nonumber \\
I^{\prime}(x, 3) &\rightarrow&
 {2^{-1/3} \pi  \over \sqrt{3}} \,
{\Gamma(5/3) \, \Gamma(1/2) \over \Gamma(11/6)} \, x^{-2/3} =
0.83 \, x^{-2/3}, \ \ \ x \gg 1.
\end{eqnarray}
For $x \gg 1$ we use the fact that most of the emission comes from
$\sin\alpha_p \simeq 1$; the integrals can then be evaluated by a
method similar to that given in Mahadevan et al. (1996).  We obtain
\begin{eqnarray}
I^{\prime}(x, 2) &\rightarrow& {\pi^{3/2} \over 4} \, x^{-1/2} \, 
	\left( 1 - {1 \over 8x^{1/2}} \right) \, \exp\left( - 2 x^{1/2} \right) 
\simeq  
1.39 \, x^{-1/4} \, \exp\left( - 2 x^{1/2} \right), \ \ \ x \gg 1, \nonumber \\
I^{\prime}(x, 3) &\rightarrow& {\pi^{3/2} \over 4} \, \exp\left( - 2 x^{1/2} \right) = 
1.39 \, \exp\left( - 2 x^{1/2} \right), \ \ \ x \gg 1.
\end{eqnarray}
\section{Compression factor in ADAFs} 
As discussed is \S4.3, the quantity $p^3 \rho^{-1}$ is conserved in
the adiabatic compression of a monatomic ideal gas, which leads to the
transformation $p \rightarrow ap \propto \rho^{1/3} p$ for the
particle momenta.  When the internal energy and pressure of the gas
contain a contribution from the magnetic field, as in ADAF models, the
form of the adiabatic invariant is more complicated.  We make the {\em
assumption} that the magnetic field always contributes a (constant)
fraction $\beta_{\rm adv}$ of the total pressure and use this to determine
the expression for the compression factor, $a$.  

The gas pressure, $P_g$, and internal energy per unit mass, $U_g$,
can be expressed in terms of an isotropic distribution function
as
\beq
P_g = {4 \pi \rho \over 3} \int dp {n(p) p^4 \over (p^2 + 1)^{1/2}}
\label{pg}
\eeq
and 
\beq
U_g = {4 \pi } \int dp n(p) p^2 \left[(p^2+1)^{1/2} - 1\right]
\equiv {b P_g/\rho},
\label{Ug}
\eeq
where, for simplicity, we set $m = c = 1$ in this Appendix.  For the
present purposes the distribution function is given by equation
(\ref{cnrmb}), that is, 
\beq
n(p) = {1 \over  (2 \pi {\tilde a}^2)^{3/2}} \exp\left[{-p^2 \over 2 {\tilde a}}\right], 
\label{df}
\eeq
where ${\tilde a} \equiv a \sqrt\theta_e$.  We treat this distribution
function and the thermodynamic quantities $U_g$ and $P_g$ as functions
of $\tilde a$ and $\rho$.  In this interpretation, the temperature
($\theta_e$) which appears in the definition of $\tilde a$ is a
constant, namely the gas temperature when the adiabatic compression
began.

For a gas in an isotropically tangled magnetic field, the total
pressure is
\beq
P_{tot} = P_g + {B^2 \over 24 \pi} \equiv P_g/\beta_{\rm adv}
\label{Pt}
\eeq
and the total internal energy per unit mass is
\beq
U_{tot} =  U_g + {B^2 \over 8 \pi \rho} = {P_{tot}
\over \rho } \left[{3 + \beta_{\rm adv}(b - 3)}\right].
\label{Ut}
\eeq
Adiabaticity requires that $dU_{tot} = P_{tot}d\ln\rho/\rho$, which
yields the following ordinary differential equation for the evolution
of $\tilde a$ as the gas is compressed.
\beq
{d \ln \tilde a \over d \ln \rho} \left[(3 + \beta_{\rm adv}(b - 3))
{d \ln P_g \over d \ln \tilde a} + \beta_{\rm adv} {d b \over d \ln
\tilde a}\right] = 1.
\label{arho}
\eeq
In the non-relativistic limit, i.e., $\tilde a \ll 1$, $b = 3/2$,
$db/d \ln \tilde a = 0$, and $d \ln P_g/ d \ln \tilde a = 2$, which
implies that $\tilde a \propto \rho^{1/(6 - 3 \beta_{\rm adv})}$.
For equipartition magnetic fields, $\tilde a \propto \rho^{2/9} \propto r^{-1/3}$.
In the relativistic limit, $\tilde a
\gg 1$, $b = 3$, $db/d \ln \tilde a = 0$, and $d \ln P_g/ d \ln \tilde a = 1$;
thus $\tilde a \propto \rho^{1/3} \propto r^{-1/2}$, independent of
$\beta_{\ rm adv}$.  We emphasize that in this context, relativistic
vs. non-relativistic refers to the magnitude of $\tilde a$. It does
not refer to the form of the distribution function, which is the same
(eq. [\ref{df}]) regardless of the particle energies.  

The above expressions for $\tilde a (\rho)$ in the non-relativistic
and relativistic limits are precise analogs of the expressions for
$T(\rho)$ for a thermal distribution given in $\S 4.2$.  This is
because the form of the distribution function (eq. [\ref{df}]) is only
relevant when calculating the evolution of $\tilde a$ through the
non-relativistic/relativistic transition (which must be done
numerically).
\section{Departures from the Adiabatically Compressed Distribution Function}
For a nearly thermal gas, the systematic and stochastic Fokker-Planck
coefficients given in \S 3 yield the timescale on which a particle's
energy changes and thus the timescale on which the distribution
becomes thermal.  For non-thermal distribution functions,
however, the distribution function may change on a timescale short
compared to the thermalization timescale.  As we now show, this
implies that the adiabatically compressed distribution function given
by equation (\ref{cnrmb}) is only strictly valid for smaller $\dot m$ than
would be inferred from the thermalization timescales given in \S 3.

We consider only self-absorbed synchrotron radiation since it is more
efficient than Coulomb collisions in modifying the distribution
function.  We also assume that, even for the non-thermal distribution
function, the Fokker-Planck coefficients are given by (\S 3.2)
\beq
{d \gamma \over dt} = C_1 B^2 \gamma^2\left(1 - {4 \langle \gamma \rangle
\over 3 \gamma}\right)
\eeq
and
\beq
{d (\Delta \gamma)^2 \over dt} = 2 C_1 B^2 { \langle \gamma \rangle
\gamma^2 \over 3},
\eeq
where we have replaced the temperature of the plasma with the average
$\gamma$ of the electrons, which is given by equation (\ref{mg}) for
the adiabatically compressed distribution function.

The timescale on which the number of particles, $n(\gamma)$, at a given $\gamma$
e-folds is given by $t_n \approx 1/|d\ln n(\gamma)/dt|$.  This, not the
systematic or diffusion timescales introduced in \S3, is the relevant
timescale when considering {\em departures from} a non-thermal
distribution function (rather than {\em approaches to} a thermal
distribution).  Evaluating $t_n$ for the adiabatically
compressed distribution function, using the Fokker--Planck equation, 
yields (cf. eqs. [\ref{fpeq}], [\ref{cnrmb}])
\beq
t_n = {1 \over C_1 B^2 \langle \gamma \rangle} \left|8 -4{\gamma
\over \langle \gamma \rangle} 
- {104 \gamma^2
\over 3 \pi {\langle \gamma \rangle}^2} 
+ {8 \gamma^3 \over \pi 
{\langle \gamma \rangle}^3} 
+ 
{64 \gamma^4  \over 3 \pi^2
{\langle \gamma \rangle}^4}\right|^{-1},
\eeq
where we have assumed that the electrons are relativistic.  For large
$\gamma$, the timescale on which the distribution function changes is
$\sim (\gamma/\langle \gamma \rangle)^4$ times faster than the usual
diffusion time, $t^{\rm s}_{\rm diff}$, and $\sim (\gamma/\langle
\gamma \rangle)^3$ times faster than the usual systematic timescale,
$t^{\rm s}_{\rm sys}$ (eq. [\ref{synchtimes}]).  The physical reason
for this is that diffusion can be much more efficient than is indicated 
by the stochastic Fokker-Planck coefficient if the non-thermal
distribution function varies strongly with energy.

For accretion times less than the minimum of the diffusion and
systematic timescales ($t_a < {\rm min}[t^{\rm s}_{\rm diff},t^{\rm
s}_{\rm sys}$]) it is incorrect to assume a thermal
distribution function, but it is only if $t_a < t_n$ that it is
strictly valid to assume that adiabatic compression yields the
distribution function given by equation (\ref{cnrmb}).  This occurs
provided that
\beq 
\dot m \lsim {7.9 \times 10^{-5} \over \langle \gamma \rangle} r \alpha^2 
\left({1 - \beta_{\rm adv} \over 0.5}\right)^{-1} 
\left|8 -4{\gamma
\over \langle \gamma \rangle} 
- {104 \gamma^2
\over 3 \pi {\langle \gamma \rangle}^2} 
+ {8 \gamma^3 \over \pi 
{\langle \gamma \rangle}^3} 
+ {64 \gamma^4  \over 3 \pi^2
{\langle \gamma \rangle}^4}\right|^{-1}
\label{mdac}
\eeq
For $\gamma \approx \langle \gamma \rangle$ there are no strong
gradients in the distribution function and so the distribution
function changes only on the energy diffusion time, i.e., $t_n \approx
t^{\rm s}_{\rm diff}$.  However, for $\gamma \approx 3 \langle \gamma \rangle$, 
which is the Lorentz factor of the electrons responsible for most of the 
synchrotron emission,  equation
(\ref{mdac}) shows that the electrons have an adiabatically compressed distribution
function only for accretion rates $\lsim 10^{-6} \alpha^2 r$. 
This is a substantially more stringent requirement than the condition 
that thermalization be inefficient.
\end{appendix}
\newpage
{
\footnotesize
\StartRef
\noindent {\large \bf References} \\
\Ref Abramowicz, M., Chen, X., Kato, S., Lasota, J. P, \& Regev, O., 1995,
ApJ, 438, L37 \\
\Ref Abramowicz, M., Czerny, B., Laso, J. P., \& Szuszkiewicz, 1988, ApJ, 332,
 	646 \\	
\Ref Balbus, S. A., \& Hawley, J.F. 1991, ApJ, 376, 214 \\
\Ref Crane, P. C., Dickel, J. R., \& Cowan, J. J., 1992, ApJ, 390, L9-L12 \\
\Ref Dermer, C. D., \& Liang, E. P.,  1989, ApJ, 339, 512-528 \\
\Ref Esin, A., 1997, ApJ, 482, in press\\
\Ref Fabian, A. C., \& Canizares, C. R., 1988, Nature, 333, 829 \\
\Ref Fabian, A. C., \& Rees, M. J., 1995, MNRAS, 277, L55-L58 \\
\Ref Frank, J., King, A., \& Raine, D., 1992, Accretion Power in
Astrophysics (Cambridge: Cambridge Univ. Press) \\
\Ref Ghisellini, G., Guilbert, P. W., \& Svensson, R., 1988, ApJ, 334, L5-L8 \\
\Ref Ghisellini, G., \& Svensson, R., 1990, in Physical Processes in 
	Hot Cosmic Plasmas, Vol. 305, ed. W. Brinkmann, A. C. Fabian, 
		\& F. Giovannelli (Dordrecht: Kluwer), 395 \\
\Ref Ghisellini, G., \& Svensson, R., 1991, MNRAS, 252, 313-318 \\
\Ref Goodman, J., \& Lee, H. M., 1989, ApJ, 337, 84-90 \\ 
\Ref Heckler, A. F., \& Kolb, E. W., 1996, ApJ, 472, L85-L88 \\
\Ref Ipser, J. R., \& Price, R. H., 1977, ApJ, 216, 578-590 \\
\Ref Ipser, J. R., \& Price, R. H., 1982, ApJ, 255, 654-673 \\
\Ref Kormendy, J., \& Richstone, D., 1995, ARA\&A, 33, 581 \\
\Ref Landau, L. D., \& Lifshitz, E. M., 1976, Mechanics, 
3rd Ed., (Oxford: Pergamon Press) \\
\Ref Lasota, J. P., Abramowicz, M. A., Chen, X., Krolik, J.,
                Narayan, R., \& Yi, I. 1996, ApJ, 462, 142\\
\Ref Mahadevan, R., Narayan, R., \& Yi., I., 1996, ApJ, 465, 327-337 \\
\Ref Mahadevan, R., 1997, ApJ, 477, 585-601 \\
\Ref Mahadevan, R., Narayan, R., \& Krolik, J., 1997, ApJ, 486, in press \\
\Ref McCray, R., 1969, ApJ, 156, 329-339 \\
\Ref Merck, M., et al., 1996, A\&A Sup., 120, 465-469 \\
\Ref Nakamura, K. E., Masaaki, K., Matsumoto, R., \& Kato, S., 1997, PASJ, in press \\
\Ref Narayan, R., \& Yi, I., 1994, ApJ, 428, L13 \\
\Ref Narayan, R., \& Yi, I., 1995a, ApJ, 444, 231 \\
\Ref Narayan, R., \& Yi, I., 1995b, ApJ, 452, 710-735 \\
\Ref Narayan, R., Yi, I., \& Mahadevan, R., 1995, Nature, 374, 623-625 \\
\Ref Narayan, R., 1996, ApJ, 462, 136 \\
\Ref Narayan, R., McClintock, J. E., \& Yi, I., 1996, ApJ, 457, 821-833 \\
\Ref Narayan, R., Barret, D., \& McClintock, 1997, ApJ, in press \\
\Ref Narayan, R., Kato, S., Honma, F., 1997, ApJ, 476, 49-60 \\
\Ref Nayakshin, S., \& Melia, F., 1997, ApJ, submitted, (astro-ph 9705011) \\
\Ref Pacholczyk, A. G., 1970, Radio Astrophysics (San Francisco: Freeman)\\
\Ref Rees, M. J., Begelman, M. C., Blandford, R. D., \& Phinney, E. S.,
1982, Nature, 295, 17 \\
\Ref Reynolds, C. S., Di Matteo, T., Fabian, A. C., Hwang, U., \&
Canizares, C. R., 1997, MNRAS, 1996, 283L, 111 \\
\Ref Rybicki, G., \& Lightman, A., 1979, Radiative Processes in
	Astrophysics (New York: John Wiley \& Sons, Inc.) \\
\Ref Shakura, N. I., \& Sunyaev, R. A., 1973, A\&A, 24, 337 \\
\Ref Shapiro, S. L., Lightman, A. P., \& Eardley, D. M. 1976, ApJ, 204,
 187 \\
\Ref Spitzer, L. Jr., 1962, Physics of Fully Ionized Gases, 2nd Ed., (New York: 
		John Wiley \& Sons, Inc.) \\
\Ref Stepney, S., \& Gilbert, P.W. 1983, MNRAS, 204, 1269 \\
\Ref Trinchieri, G., Fabbiano, G., Canizares, C. R., 1986, ApJ, 310,
	637-659 \\
 
}
\newpage
\newpage
\noindent{\bf Figure Captions.} \\

\noindent Figure 1:  For accretion rates above a given curve in the figure, 
an electron with Lorentz factor $\gamma$ is able to thermalize with a
background plasma of temperature $\theta_e$.  The dashed lines
represent thermalization through Coulomb collisions, while the solid
lines represent thermalization through synchrotron self--absorption.
The curves are determined by whether the dominant thermalizing process
is systematic or diffusive acceleration (cf. eqs. [\ref{eemdot}],
[\ref{synchmdot}]).  While thermalization by Coulomb collisions is,
for a fixed temperature, independent of the radius $r$
(cf. eq.[\ref{eemdot}]), the accretion rates required for
thermalization by synchrotron self--absorption increase linearly with
$r$ (cf. eq. [\ref{synchmdot}]).

\noindent Figure 2:  a) A comparison of the momentum distribution function 
for a thermal gas of temperature $\theta_e \sim 1$ with that of an
adiabatically compressed collisionless gas of the same energy. The adiabatically
compressed gas is initially thermal and non--relativistic, but is then
compressed to relativistic energies, resulting in a non--thermal
distribution function.  b) Analogous to a), but for the adiabatic
expansion of a gas from relativistic to non--relativistic energies.

\noindent Figure 3:  a) A comparison of the synchrotron radiation from 
a sphere of radius $\simeq 7 \times 10^{11}$ cm for an adiabatically
compressed (solid line) and thermal (dashed line) gas with the same
average energy.  The electron number density and magnetic field strength
are constant throughout the sphere ($n_e \simeq 10^{10}$ cm$^{-3}$ and  
$B\simeq 200$ Gauss).  At
low frequencies the radiation is highly self--absorbed, and the
observed intensity is given by the source function, not the emissivity
(dotted line) of the respective gases.  b) The total synchrotron
spectrum from an ADAF ($m = 2\times 10^6$, $\dot{m} = 10^{-5}$) for
adiabatically compressed electrons (solid line) and thermal electrons
(dashed line) of the same average energy, which is assumed to be
constant throughout the ADAF.  This spectrum is obtained by summing
the individual spectrum from each radius in the accretion flow, and
thus accounts for the variation in electron number density and
magnetic field strength with radius.
\newpage
\begin{figure}
\epsffile{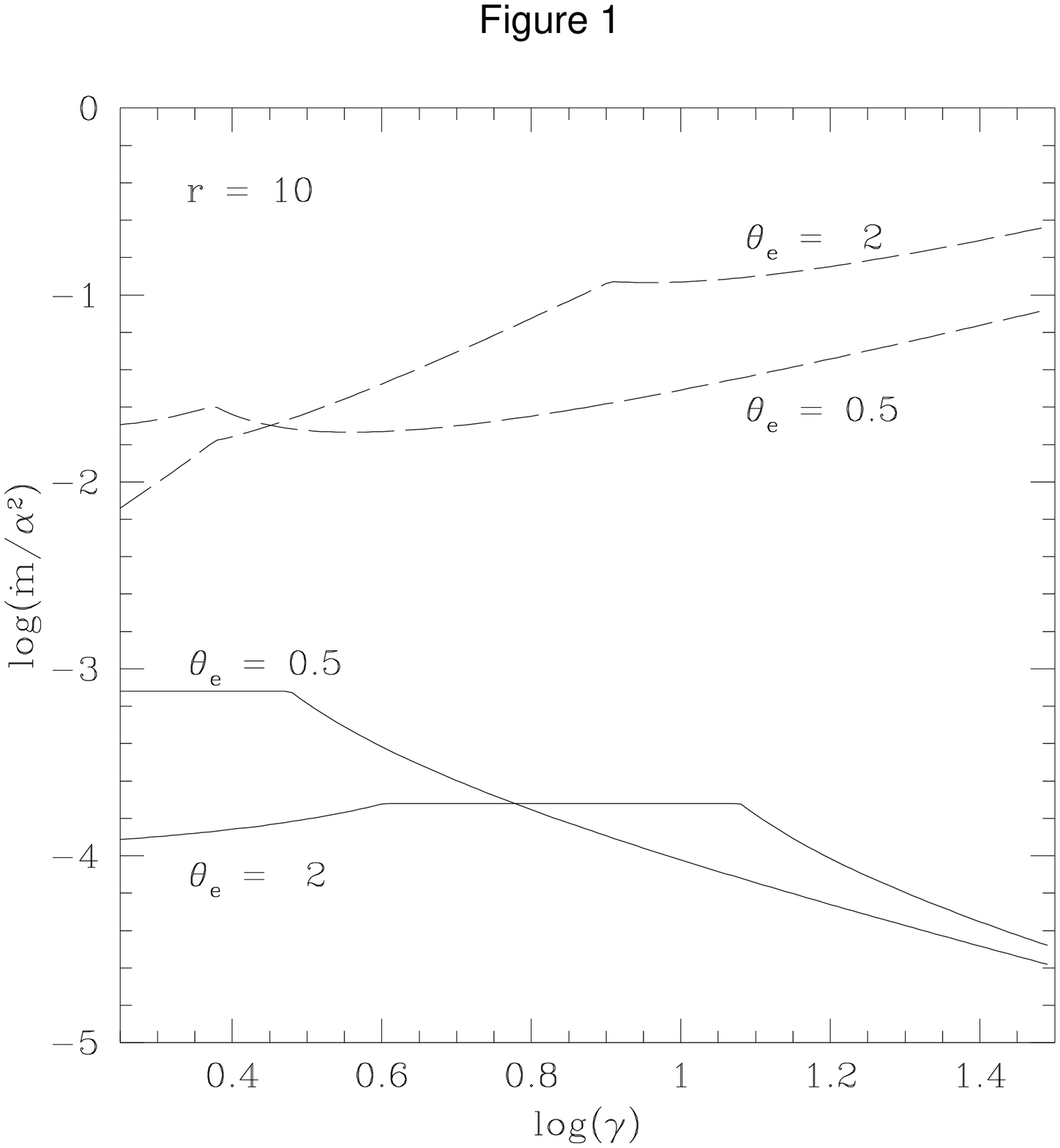}
\end{figure}
\begin{figure}
\epsffile{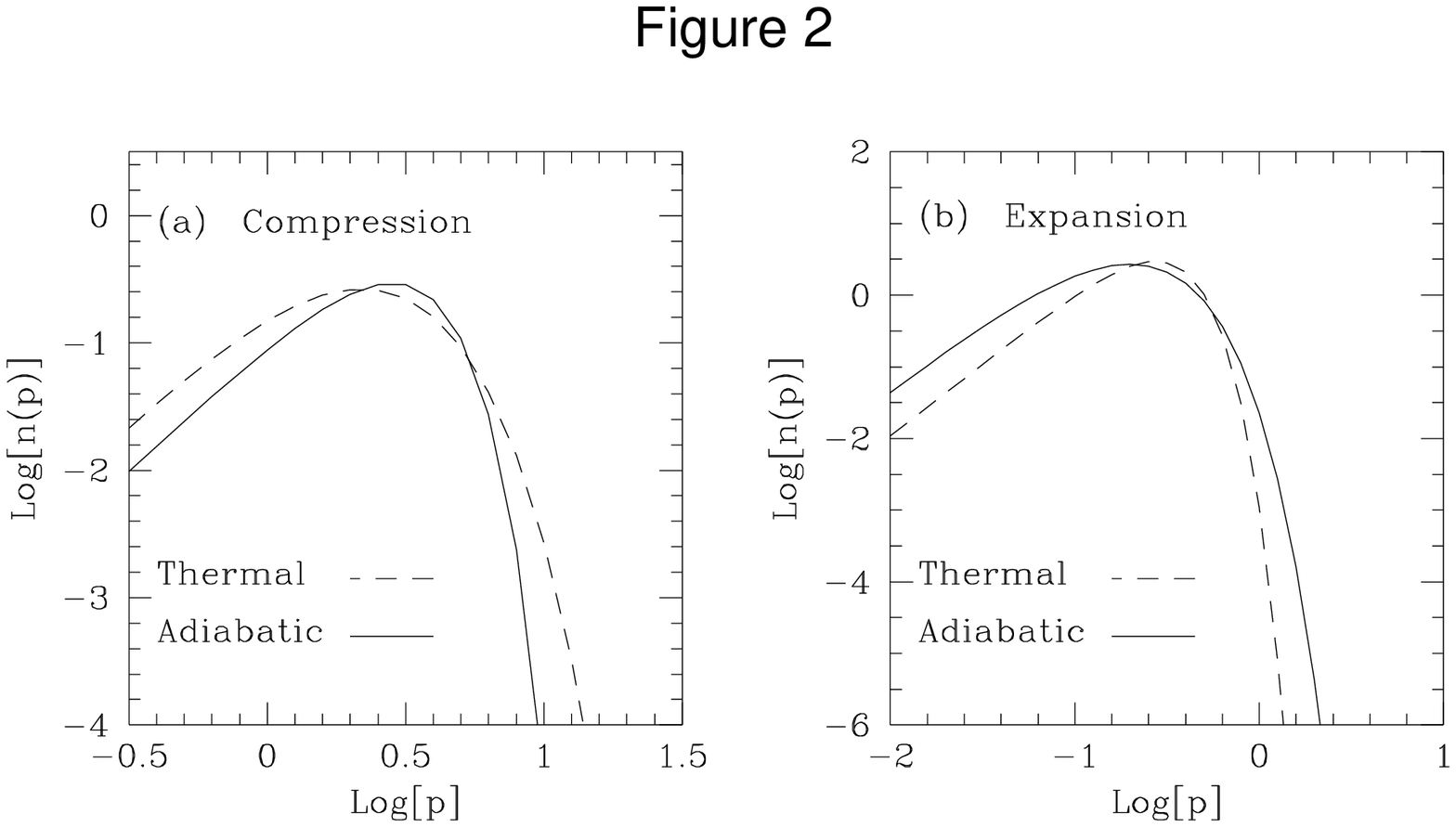}
\end{figure}
\begin{figure}
\epsffile{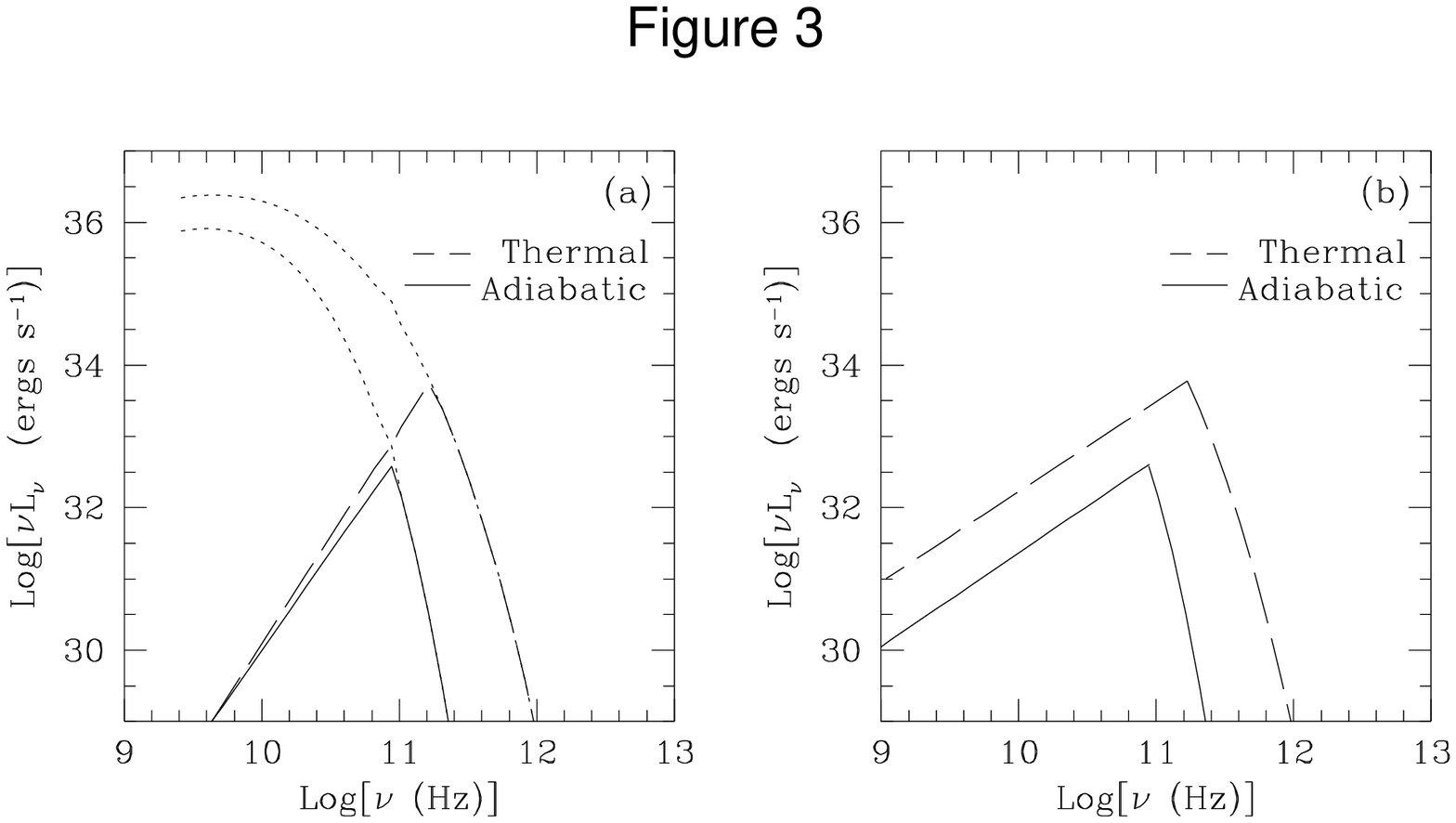}
\end{figure}
\end{document}